\documentclass{article}
\usepackage[utf8]{inputenc}
\usepackage{eqnarray,amsmath}

\newenvironment{prof}[1][Proof]{\noindent\textbf{#1.} }{\ \rule{0.5em}{0.5em}}

\newtheorem{lemma}{Lemma}

\newcommand\pernb[1]{n_{#1}}
\newcommand\start[1]{s_{#1}}
\newcommand\horis[1]{u_{#1}}
\newcommand\vertis[1]{v_{#1}}
\newcommand\horip[1]{x_{#1}}
\newcommand\vertip[1]{y_{#1}}
\newcommand\larg{w}
\newcommand\hyperp{H}
\newcommand\bigrect{\mathcal{R}}
\newcommand\rect[1]{R_{#1}}
\newcommand\height[1]{h_{#1}}
\newcommand\base[1]{b_{#1}}
\newcommand\job[1]{J_{#1}}
\newcommand\bbase[1]{B_{#1}}
\newcommand\proctime[1]{p_{#1}}
\newcommand\per[1]{T_{#1}}
\newcommand\pertask[1]{\per{\pernb{#1}}}

\newcommand\flip{flip}
\newcommand\bflip{Bf}
\title{Periodic Scheduling and Packing Problems}
\author{Claire Hanen, Zdenek Hanzalek }

\begin{document}

\maketitle

\section{Introduction}

An embedded processor that executes computation tasks needed for control loops is a typical example of an application that must operate periodically within the bounds of the control loops periods. Periodic scheduling problems are frequent in a many applications, including avionics~\cite{beji2014smt},  automotive~\cite{minaeva2017time}, software-defined radio~\cite{Bar-Noy2002}, and periodic machine maintenance~\cite{wei1983periodic}.

This paper studies relations between periodic scheduling problem and packing problem. Namely it shows equivalence of harmonic periodic scheduling problem and  ruled  harmonic  2D  packing problem. 

\section{State of the art}

Some basic preemptive periodic scheduling problems are pseudo-polynomial ~\cite{baruah1990algorithms}, while a non-preemptive periodic scheduling problem is strongly NP-hard~\cite{jeffay1991non} in most of the cases. In this paper we deal with non-preemptive and zero-jitter periodic scheduling. 

On a single resource, \cite{jeffay1991non} showed that the non-preemptive periodic scheduling with arbitrary task initial phases (i.e., each task is released at its release-date and must be finished within its period time units) and no criterion is strongly NP-hard. 
For the case of the harmonic period set the problem seems to be an easier, since there are efficient heuristics algorithms (\cite{dvorak2014multi}).
However, it is known by Cai \cite{cai1996nonpreemptive} that $1|T_i^{harm},p_i^{nonharm}|-$ is NP complete in the strong sense, while periodic scheduling on parallel identical resources $P|\{T_i,p_i\}^{harm}|-$ is known to be polynomial by Korst \cite{korst1996scheduling}.

Periodic scheduling on one machine with unit processing times was shown to be NP-complete by~\cite{Bar-Noy2002} using the reduction from the graph coloring problem. Furthermore, \cite{jacobs2014new} proved that the problem is strongly NP-hard. 


Relations between 1 machine case and 2D packing has been noticed in Korst \cite{korst1996scheduling} and Zhao as remarks, and formalized for powers of 2 periods in Lukasiewicz \cite{lukasiewycz2009flexray}.

\section{2DPacking and the 1 machine harmonic case}
\subsection{Problem definition}
In this section we consider a set of $n$ independent jobs, $\job{1},\ldots,\job{n}$. Each job $\job{i}$ is characterized  by its processing time $\proctime{i}$, and its period $\pertask{i}$. Periods  belong to an harmonic set $\per{1},\ldots,\per{r}$, so that $\forall k\in\{2,\ldots,r\}, \per{k}=\base{k}\per{k-1}$ where $\base{k}$ is an integer. We also  consider a value $\larg$ such that $\per{1}=\base{1}\larg$, and assume that any processing time of a job is not greater than $\larg$.

We denote by $\bbase{k}=\base{1}\times\ldots \times \base{k}$. Notice that
\begin{equation}
    \per{k}=\bbase{k}\larg
\end{equation}
so that for any job $\job{i}$, 
\begin{equation}
    \pertask{i}=\bbase{\pernb{i}}\larg
\end{equation}
Let $H=\bbase{r}$. 

Any periodic schedule defines for each job $\job{i}$ a starting time $\start{i}\le \pertask{i}$ of its first occurrence. An occurrence of $i$ will start  at each time  $\start{i}+k\pertask{i}, k\ge 0$.
In this section we assume that the jobs are to be performed on a single machine.

Clearly, due to the periodicity of the jobs, if the periodic schedule has no resource conflict in the time interval $[0,wH]$, it has no conflict. 

We define $$\height{i}=\frac{\larg\hyperp}{\pertask{i}}$$  the number of values of  $k$ such that $k\ge 0, \start{i}+k\pertask{i}<\larg\hyperp$. 

 The question whether there exists a 1 machine schedule such that no collision occur  can be formulated as follows:
for any integer $k,k'$ with $0\le k< \height{i}$ and  $0\le k'<\height{j}$ and for any two jobs $\job{i},\job{j}$ one of the two following conditions hold:
\begin{align}
  \start{i}+k\pertask{i}+\proctime{i}&\le  \start{j}+k'\pertask{j}\\
     \start{j}+k'\pertask{j}+\proctime{j}&\le  \start{i}+k\pertask{i}
\end{align}

Now, we can decompose the starting times along with the harmonic periods as follows: $$\forall \job{i},\quad \start{i}=\horis{i}+\vertis{i}\larg$$ where $\horis{i}< \larg$. Suppose that $\larg$ is defined such that $\horis{i}+\proctime{i}\le \larg$ in each considered feasible schedule. (This property is always true if we choose $\larg =\per{1}$, but as we will see, different values might be interesting to choose).
Let us notice that $0\le \vertis{i}< \bbase{\pernb{i}}$ (i.e.$ \vertis{i}<\frac{\pertask{i}}{w}\le H$.

Notice that as job $\job{i}$ has period $\pertask{i}$, then an occurrence of $i$ will start  at each time  $$ \start{i}+k\pertask{i} =\horis{i}+ (\vertis{i}+k\bbase{\pernb{i}})\larg $$

We can thus define the necessary and sufficient condition of a collision to occur in a periodic schedule:
\begin{lemma}
A periodic schedule induces a collision between two jobs $\job{i},\job{j}$ such that $\pernb{i}\le\pernb{j}$ if and only if the two following conditions hold:
\begin{equation}
 \horis{i}<\horis{j}+\proctime{j}\quad and\quad  \horis{j}<\horis{i}+\proctime{i}
    \label{collisionhoriz}
\end{equation}
\begin{equation}
\exists k,\quad \vertis{j}=\vertis{i}+k\bbase{\pernb{i}}
    \label{collisionvert}
\end{equation}
\end{lemma}

\subsection{Mixed radix system and flip transformation}
Let $y$ be any integer. It is known that $y< H$ can be decomposed uniquely according to the mixed radix numerical system $\base{}=(\base{1},\ldots,\base{r})$ as follows:
$$y=y_1+y_2\bbase{1}+ y_3\bbase{2}+\ldots+y_r\bbase{r-1}$$
with $\forall i> 0, y_i<\base{i}$. We denote this decomposition as follows \begin{equation}
    [y]_{\base{}}=[y_1]_{\base{1}}\ldots [y_r]_{\base{r}}
\end{equation}

Notice that in the usual base decomposition (for example base 2 for binary decomposition), the components of the base vector $\base{}$ are all equal (to 2 for binary decomposition).

As the partial products depend on the base vector, we introduce the vector in the notation, since in the following the vector may change: $$\bbase{j}(\base{})=(\base{1},\ldots,\base{j})$$

Let us generalize the transformation proposed by Lukasievicz for the binary decomposition of a number by defining two operators:
Let $\bflip{(\base{},k)}$ be the operator that flips the k first component of the base vector $\base{}$:
\begin{equation}
    \bflip{(\base{},k)}=(\base{k},\base{k-1},\ldots,\base{1},\base{k+1}\ldots,\base{r})\label{def:bflip}
\end{equation}

Observe that \begin{align}\bflip{(\bflip{(b,k)},k)}=b\\
\bbase{k}(\bflip(b,k))=\bbase{k}(b)\end{align}

Let now $\flip(y,k,\base{})$ be the number constructed by flipping the $k$ first digits of the decomposition $[y]_{\base{}}$ to give a number expressed with respect to base vector $\base{}'=\bflip{(\base,k)}$:
\begin{equation}
[\flip(y,k,\base{})]_{\base{}'}=[y_k]_{\base{k}}\ldots [y_1]_{\base{1}}[y_{k+1}]_{\base{k+1}}\ldots[y_r]_{\base{r}}
    \label{def:flip}
\end{equation}

Hence we have:
\begin{equation}
    \flip(y,k,\base{})=y_k+y_{k-1}\base{k}+\ldots+y_1(\base{2}\ldots\base{k})+y_{k+1}(\base{1}\ldots\base{k})+\ldots+y_r(\base{1}\ldots\base{r-1})\end{equation}

This flip operation has some important properties that will be used to transform the scheduling problem into an equivalent packing problem.

\begin{lemma}
for any integer $y<H$, \begin{equation}\flip{(\flip{(y,k,b)},k,\bflip{(b,k)})}=y\end{equation}

 if $[y_k]_{\base{k}}<\base{k}-1$, then \begin{equation}\flip{(y+\bbase{k-1}(b)),k,b)}=\flip{(y,k,b)}+1\end{equation}
 and if $[y_k]_{\base{k}}>0$,
 \begin{equation}\flip{(x-\bbase{k-1}(b),k,b)}=\flip{(y,k,b)}-1\end{equation}
\label{lemma:flip}
\end{lemma}
This implies that equidistant integers (with distance $\bbase{k-1}(b)$) after the flip transformation, become consecutive integers.

\subsection{Periodic scheduling and ruled harmonic 2D packing definition}
Now consider the following 2D packing problem associated with the original scheduling problem. We are given a rectangle $\bigrect$ of length $\larg$ and height $\height{}$. For each job $\job{i}$ we define a rectangle $\rect{i}$ of length $\proctime{i}$ and height $\height{i}$.

Assume that we want to pack rectangles $\rect{i}$ into the big rectangle $\bigrect$ Hence we have to define a position of the lowest left point of each rectangle, with coordinates $(\horip{i},\vertip{i})$ so that no collision occurs. Moreover we assume that in the packings we consider, we must have the additional property:

\begin{equation}
    \forall i, \vertip{i}\%\height{i}=0
    \label{eq:addpropertypack}
\end{equation}

\begin{lemma} In a packing satisfying property (\ref{eq:addpropertypack}) A collision  between rectangles $\rect{i}$ and $\rect{j}$  such that $\height{j}\le\height{i}$ occurs iff the two following condition hold:
\begin{equation}
 \vertip{i}\le \vertip{j}<\vertip{i}+\height{i}, 
   \label{eq:collisionpacking}
 \end{equation}
 \begin{equation}\
  \horip{j}<\horip{i}+\proctime{i},\quad \horip{i}<\horip{j}+\proctime{j}
\end{equation}
\label{lemma:collisionpacking}
\end{lemma}
\begin{prof}
obviously a collision occur if: 
\begin{align}
   \vertip{i} \le \vertip{j}<\vertip{i}+\height{i} &\quad or&  \vertip{j} \le \vertip{i}<\vertip{j}+\height{j}\label{vertipcollision}\\
 \horip{j}<\horip{i}+\proctime{i},\quad \horip{i}<\horip{j}+\proctime{j}\label{horipcollision}
\end{align}
Assuming $\height{j}\le\height{i}$,  $\height{j}$ divides $\height{i}$ and also divides $\vertip{j}$ and $\vertip{i}$ the condition $\vertip{j} \le \vertip{i}<\vertip{j}+\height{j}$ cannot hold.
\end{prof}

\subsection{Equivalence of the two problems}
In this section, we prove that the feasibility of a periodic schedule and 2D packing feasibility are equivalent problems. 
To this purpose we define, for any periodic schedule $\sigma$ (which defines associated values $s,u,v$) an associated packing as follows:
\begin{equation}
\forall \job{i},\quad \left\{\begin{array}{ll}\horip{i}^\sigma&=\horis{i}\\
\vertip{i}^\sigma&=\height{i}\flip{(\vertis{i},\pernb{i},\base{})}\end{array}\right.
\label{sched2pack}
\end{equation}
Conversely if a packing $\pi$ is given, defining $(\horip{i},\vertip{i})$ for each job $\job{i}$ then an associated schedule is defined as follows:
\begin{equation}
\forall \job{i},\quad \left\{\begin{array}{ll}\horis{i}^\pi&=\horip{i}\\
\vertis{i}^\pi&=\flip{(\frac{\vertip{i}}{\height{i}},\pernb{i},\bflip{(\base{},\pernb{i})})}\end{array}\right.
\label{pack2sched}
\end{equation}

\begin{lemma}
A feasible periodic schedule $\sigma$ on 1 machine defines a feasible 2D packing satisfying property (\ref{eq:addpropertypack}).
\label{lemma:equivsched2packing}
\end{lemma}
\begin{prof}
Let us consider the packing parameters defined by relation (\ref{sched2pack}) associated to a schedule $\sigma$.
First observe that for any job $\job{i}$, $\horip{i}^\sigma<w$, and by construction $\vertip{i}^\sigma\% \height{i}=0$, so that condition (\ref{eq:addpropertypack}) is met.
Now, assume that a collision occurs in the 2D packing between jobs $\job{i}$ and $\job{j}$ such that $\height{j}\le\height{i}$ , so that $\pertask{i}\le \pertask{j}$.
According to lemma (\ref{lemma:collisionpacking}) 
$$\vertip{i}^\sigma \le \vertip{j}^\sigma<\vertip{i}^\sigma+\height{i} 
$$
Assume that $ \vertip{j}^\sigma=\vertip{i}^\sigma+\Delta$. Notice that as $\height{j}$ divides $\height{i}$, and as $\height{j}$ divides $ \vertip{j}^\sigma$, it should also divide $\Delta$

moreover, \begin{align}
    \height{j}\flip{(\vertis{j},\pernb{j},\base{})}=   \height{i}\flip{(\vertis{i},\pernb{i},\base{})}+\Delta\\
    \flip{(\vertis{j},\pernb{j},\base{})}=\frac{\height{i}}{\height{j}}\flip{(\vertis{i},\pernb{i},\base{})}+\frac{\Delta}{\height{j}}\\
    with\ \frac{\Delta}{\height{j}}<\frac{\height{i}}{\height{j}}
\end{align}
Now according to lemma \ref{lemma:flip}, 
\begin{align}
\vertis{j}=\flip{(\frac{\vertip{j}^\sigma}{\height{j}},\pernb{j},\bflip{(\base{},\pernb{j})})}\\
=\flip{(\frac{\height{i}}{\height{j}}\flip{(\vertis{i},\pernb{i},\base{})}+\frac{\Delta}{\height{j}},\pernb{j},\bflip{(\base{},\pernb{j}}))}
\end{align}

Now the first $\pernb{j}-\pernb{i}$ digits of $\frac{\height{i}}{\height{j}}\flip{(\vertis{i},\pernb{i},\base{})}$ in the base representation $\bflip{(\base{},\pernb{j})}$ equal $0$, whereas  as $\frac{\Delta}{\height{j}} \le\frac{\height{i}}{\height{j}}$ its last $\pernb{i}$ digits are equal to $0$. So the flip operation can be applied separately on the two numbers:
\begin{align}
\vertis{j}= 
\flip{(\frac{\height{i}}{\height{j}}
\flip{(\vertis{i},\pernb{i},\base{})}
,\pernb{j},\bflip{(\base{},\pernb{j})})}
+\flip{(\frac{\Delta}{\height{j}},\pernb{j},\bflip{(\base{},\pernb{j})})}
\end{align}

Now, the first $\pernb{i}$ digits of $\flip{(\frac{\height{i}}{\height{j}}\flip{(\vertis{i},\pernb{i},\base{})},\pernb{j},\bflip{(\base{},\pernb{j}}))}$ in the base representation $\base{}$ are equal to the first $\pernb{i}$ first digits of $\vertis{i}$ in base representation $\base{}$ (and the last digits equal 0). Moreover the number $\flip{(\frac{\Delta}{\height{j}},\pernb{j},\bflip{(\base{},\pernb{j}}))}$ has its first $\pernb{i}$ digits equal to $0$ in the base representation $\base{}$. So that it is a multiple of $\bbase{i}$. 
\begin{align}
    \vertis{j}= \vertis{i}+ \delta \bbase{\pernb{i}}
\end{align}
so that condition (\ref{collisionvert}) occurs.
Now, if  condition  (\ref{horipcollision}) is met then so is condition (\ref{collisionhoriz}). 
Hence, there would be a collision in the schedule, the contradiction.
\end{prof}
\begin{lemma}
Any 2D packing satisfying property \ref{eq:addpropertypack} defines a feasible periodic schedule.
\label{lemma:equivpacking2sched}
\end{lemma}
\begin{prof}
Let us consider a feasible packing $\pi$, and assume that there is a collision between two jobs $\job{i}$ and $\job{j}$ with  $\pertask{i}\le \pertask{j}$.
Notice that if condition (\ref{collisionhoriz}) is satisfied then so is condition (\ref{horipcollision}).
Now assume condition (\ref{collisionvert}) is satisfied, so that we have 
\begin{align}
   \vertis{j}^\pi= \vertis{i}^\pi+ \delta \bbase{\pernb{i}}
\end{align}

So according to lemma \ref{lemma:flip} and the reversibility of the base flip ($\bflip{(\bflip(b,k),k)}=b$),

$$\vertip{j}=\height{j}\flip{(\vertis{j}^\pi,\pernb{j},b)}$$
So,
\begin{align}
\frac{\vertip{j}}{\height{j}}=\flip{(\vertis{i}^\pi+ \delta \bbase{\pernb{i}},\pernb{j},b)}\\
=\flip{(\flip{(\frac{\vertip{i}}{\height{i}},\pernb{i},\bflip{(b,\pernb{i})})}+ \delta \bbase{\pernb{i}},\pernb{j},b)}\end{align}

Notice that as  $\vertip{i}< H$ so that $\frac{\vertip{i}}{\height{i}}<\bbase{\pernb{i}}(b)=\bbase{\pernb{i}}(\bflip{(b,\pernb{i})})$ and thus  when performing the operation $\flip{(\frac{\vertip{i}}{\height{i}},\pernb{i},\bflip{(b,\pernb{i})})}$, we get a number still less than $\bbase{(\pernb{i})}$ which has non null positions only in the $\pernb{i}$ first digits in base vector $b$.

So according to lemma \ref{lemma:flip}, 
\begin{align*}
    \flip{(\flip{(\frac{\vertip{i}}{\height{i}},\pernb{i},\bflip{(b,\pernb{i})})}+ \delta \bbase{\pernb{i}}(b),\pernb{j},b)}\\=\flip{(\flip{(\frac{\vertip{i}}{\height{i}},\pernb{i},\bflip{(b,\pernb{i})})},\pernb{j},b)} + \delta\\
    =\frac{\bbase{\pernb{j}}(b)}{\bbase{\pernb{i}}(b)}\flip{(\flip{(\frac{\vertip{i}}{\height{i}},\pernb{i},\bflip{(b,\pernb{i})})},\pernb{i},b)}+\delta
    \end{align*}
   Hence \begin{align}
   \frac{\vertip{j}}{\height{j}} =\frac{\bbase{\pernb{j}}(b)}{\bbase{\pernb{i}}(b)}\times\frac{\vertip{i}}{\height{i}}+\delta
\end{align}

Now $\height{i}\bbase{\pernb{i}}(b)=\height{j}\bbase{\pernb{j}}(b)=H$, so that 
\begin{equation}
    \vertip{j}=\vertip{i}+\delta
\end{equation}
Now $\delta<\height{i}$ since $\vertis{j}^\pi= \vertis{i}^\pi+ \delta \bbase{\pernb{i}}<H=\height{i}\bbase{\pernb{i}}$. So condition (\ref{vertipcollision}) occurs, a contradiction.
\end{prof}
\subsection{Approximation}
Now we can look at the problem of minimizing $w$ such that a schedule exists. This might be interesting while considering for example a part of the cycle that is booked for time triggered traffic, while the rest is booked for event triggered communications.
As done for example in the paper of Zhao, Qin and liu 
We can use our 2D packing transformation, and try to adapt approximation algorithms for the underlying
\begin{lemma}
The FFDH algorithm from Coffman Garey, Johnson algorithm for 2D strip packing with harmonic length produce packings that can be modified to satisfy property \ref{eq:addpropertypack}.
\label{lemma:CGJ}
\end{lemma}
\begin{prof}
The  FFDH algorithm basically sorts the rectangles according to their length (in non increasing order) and pack them by "shelves" or strips. A shelve is open when putting a rectangle at its base. The other rectangles (with smaller length) are then placed one upon the other until the total height is reached (or no more rectangle is to be placed). Then a new shelve is openened.

We claim that in each shelve or strip, a reordering can be done on the rectangles so that their coordinates $(\horip{i},\vertip{i})$ satisfy the constraint \ref{eq:addpropertypack}, due to the harmonic nature of the periods.
Consider a shelve, in which rectangles $\rect{i_1},\ldots,\rect{i_k}$ are stored in this order (so that $\proctime{i_1}\ge\ldots\ge\proctime{i_k}$.
Notice first that any permutation of rectangles on the shelve lead to a feasible packing.
So we can sort them by non increasing height. And we can then prove that the vertical coordinate $\vertip{i}$ of these rectangles satisfy constraint \ref{eq:addpropertypack}.
\end{prof}

\begin{lemma}
If we apply modified FFDH algorithm to 2D strip packing, in order to minimize $w$, we define a periodic schedule such that \begin{equation}w^{CGJ}\le w^{opt}+max_ip_i\end{equation}
\label{lemma:approx}
\end{lemma}
\subsection{parallel machine case}
The decision problem is a bin packing problem. 2D bin packing algorithm can be extended to solve with approximation the problem of computing the minimum number of machines necessary to compute a set of periodic jobs.

\subsection{Release dates and deadlines}
Assume now that each job $J_i$ has a release date $r_i$ and a deadline $d_i$, which means that the successive occurrences of $J_i$ satisfy $s_i+kT_i\ge r_i+kT_i$ and $s_i+kT_i+p_i\le d_i+kT_i$ so that $s_i\ge r_i$ and $s_i+p_i\le d_i$.

Assume that $r_i $ and $d_i$ are multiples of $w$. Then it will imply that the constraint on $s_i$ can be expressed on $\vertis{i}$:
\begin{equation}
    \frac{r_i}{w}\le \vertis{i}\le\frac{d_i}{w}
\end{equation}
This will induce a constraint on the associated packing problem
so that not all positions of the rectangle $\rect{i}$ will be available. Notice that the allowed positions are not consecutive ones.

 \section{Conclusion}
This paper has shown equivalence of the harmonic periodic scheduling problem and the ruled harmonic 2D packing problem and thus it contributed to understanding of the periodic problem complexity. 

\bibliographystyle{plain} 
\bibliography{bibliography}
\end{document}